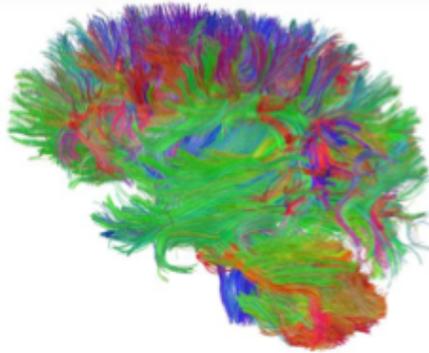

Jintao Long, Richard Tomsett, Marcus Kaiser

# Mapping neuron positions to curved cortical surfaces

Technical Report No. 3
Wednesday, 14 May 2014

Dynamic Connectome Lab
http://www.biological-networks.org/

# Mapping neuron positions to curved cortical surfaces

Jintao Long, Richard Tomsett, Marcus Kaiser

School of Computing Science, Newcastle University, UK

**ABSTRACT**

**Motivation:** Real cortical tissue curves and folds according to experiment data. However, the cortical slice model is organized in flat layers currently. This project would extend the cortical slice model in order to be able to specify arbitrary curvatures to the cortical layers and investigate the effect of curvature and folding on neuron density and layer morphology.

## 1 INTRODUCTION

The neocortex which is also named neopallium and isocortex is one of the components of the mammalian brains. It consists six layers labeled I to VI from innermost layers to outermost layers, covering the outer layer of cerebral hemispheres. As part of cerebral cortex, neocortex participates in many higher functions such as conscious thought, language, reasoning and sensory perception.

The cerebral cortex is mainly configured by neocortex. Categorized into six layers based on depth from a certain layer to the cortical surface, it contains about 10 and 14 billion neurons. The six layers are labeled with Roman numerals. Layer I is the molecular layer anchoring very few neurons. Layer II is the external granular layer. Layer III is the external pyramidal layer. Layer IV is the internal granular layer. Layer V is the internal pyramidal layer. And layer VI is the multiform or fusiform layer. Every cortical layer is consisted of various neuronal shapes, sizes and density in addition to the different organizations of nerve fibers.

Generally, neocortex is divided into four lobes that are occipital lobe, parietal lobe, frontal lobe and temporal lobes respectively according to the location of regions. In addition, neocortex can also be divided based on the practical function on particular regions. (Trappenberg, 2010)

In order to provide insight of some aspects of a complex system or hypothesis, a model as a simplification of a system is hence needed to be built to better investigate and understand the underlying mechanisms. Likewise, applying models to simplify actual regions are indispensable method in computational neuroscience to comprehend different features and characteristics on brain figures and functions.

Simplified versions of neuron models are frequently used in many neuroscience studies for less computational complexity and more explicit exhibition of target features. A simplified neuron model commonly used in network simulations is the integrate-and-fire (IF) model (Lapicque, 1907; Stein, 1967; Tuckwell, 1988), though inevitably certain limitations can be found obviously in traditional leaky IF models. First, the sub-threshold dynamic of neuron membrane potential is not efficiently approximated. Furthermore, the diversity of response pattern that can be seen in real neurons cannot be represented in the model (Trappenberg, 2010). To overcome these deficits, new generations of models need to be developed.

**Fig. 1.** Flat cortical surface and its underlying flat cortical layers.

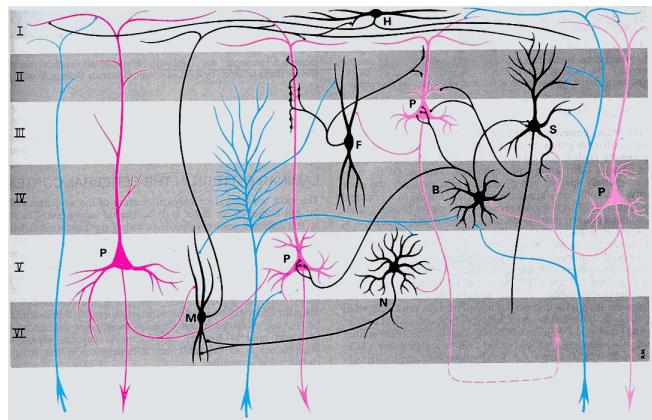

The compartmental models applied in neural transmission are invented inspired by communication cables innovated in 19th century by Lord Kelvin. Since then neuroscientists have been endeavouring to shrink the number of compartments so as to speed up the simulations. However, few compartments assigned to the model may contribute to the loss of electrotonic characteristics of the neurons. One of the models trying to solve this problem is the 'cartoon representation' which is a 24 compartment model of cortical pyramidal neuron developed by Stratford (Stratford et al, 1989). Based on this approach, Bush and Sejnowski (Bush and Sejnowski, 1993) developed a reduced compartmental model for neocortical pyramidal cells combining with some simpler methods. It is also the single neuron model that the simulation of neuron network in this study is going to apply.

Fluctuations in the postsynaptic membrane voltage are observed commonly in the in vivo neuronal experiments, due to the synchronicity of the background input. This indicates that neurons in the brain remain spike-time variability which can create disparity between our simulated model and real neuron firing environment if it is not considered in the process of model building. Generally, the spiking noise of neurons can be detected from both relatively low-frequency responses and high-frequency responses. It is worth mentioned that the impact of input fluctuation is not always clear until the discovery of sensitivity from mPFC L5 pyramidal neurons to input fluctuation (Arsiero et al, 2007). Therefore, an appropriate choice of random process should be adopted to mimic this variability.

**Fig. 2.** Curved cortical surface and its underlying curved cortical layers.

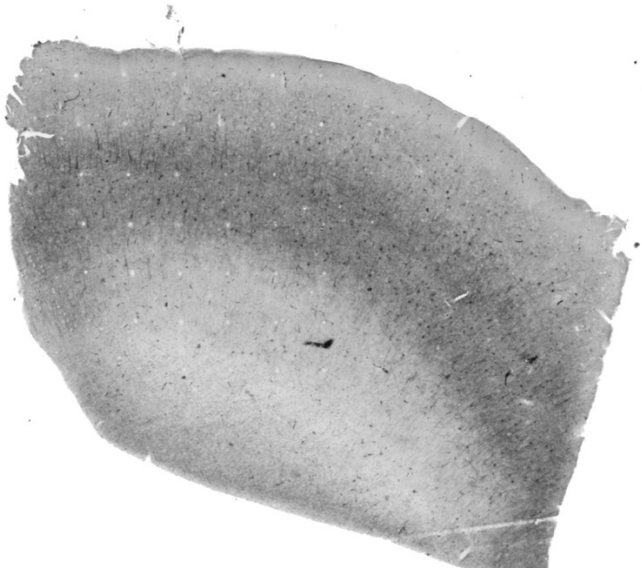

Ornstein-Uhlenbeck process (Uhlenbeck and Ornstein, 1930) which is a stochastic process is going to be used to resemble the chaotic fluctuation of input. It is generally an adjusted version of the Wiener process, which basically believes in the mean retrieval and tendency of a variable to move back to central location after a change of state. The process itself is stationary, Gaussian and Markovian, making it the only nontrivial process that can fulfill these three premises.

The cortical surface, in most cases, is not flat with constant layer boundaries but containing many curves and folds (see Fig. 2). This alternation could render series of complicated effect of many neuronal variables, such as neuronal density, synapse density and axonal arbourisation etc., which could deteriorate the accuracy and credibility of a model. Several studies have examined the impact made by folding on laminar morphology to be intense enough to greatly influence the cerebral cortical landscape and its underlying properties. Hilgetag (Hilgetag and Barbas, 2006) found the changes of both the relative and absolute layer thickness caused by the folding of layers. Moreover, it is found that there could be neuron migration occurring when the cortical layers are being bent. And the migration direction, as hypothesized and verified, could be as the same direction at which neuron moves to layers with more compression. It means that for a gyral region, the neurons tend to move to deeper layers. While an opposite case happens for neurons in a sulcal region. This study clearly demonstrates that the effect of curvature on the laminar properties of cerebral cortex.

However, previous models reconstructing the neuron network in cortical tissue neglect these properties of cortical surface (See Fig. 1; Fig. 4). Therefore, a model that takes the curvature of cortical surface into consideration is needed. This study aims to develop effective transformation methods for existing model to simulate curved cortical layers and evaluate different methods created so as to specify arbitrary curvature to the cortical layers.

Curve fitting methods are needed in the study to define the constraint boundary in Monte Carlo experiment in the calculation of the curved cortical slice area. Polynomial regression (Gergonne, 1974; Stigler, 1974) which is a form of linear regression is used as the method of curve fitting.

## 2  METHODS

The method section is basically divided into five parts – preparation, transformation, apical dendrite orientation adjustment, evaluation and real curve specification (see Fig. 3).

### 2.1  Preparation

The simulations of model composition, model size, coordinate system and topology of neuron network are adopted from Tomsett et al (Tomsett RJ). In this study, only the morphology of neuron network is under investigation. Therefore, the simulation is initiated to the phase where there is no input stimulation to the neuron network. By employing the soma position and compartment position generated by the initiation, further modifications can be made to adjust the coordinates of neurons in order to mimic the situation of a curved cortical surface.

It is assumed in this study that the coordinates describing the thickness of the cortical slice (which is y axis in the simulation) of neurons remain unchanged after transformations. Also the transformations are all restricted only to one cortical layer, which is layer V in the study.

### 2.2  Transformation

The process of transformation is basically parted into the following procedures.

*2.2.1  Splitting* The cortical slice is first horizontally split into assigned number of pieces with equal width of each piece of column. In this procedure, it is equivalent to categorizing neurons according to their x coordinates. If the x coordinate of a particular neuron locates within a certain numeric range, a column index specified for this range is assigned for that neuron.

*2.2.2  Methods for transformation* Adjustment of neurons on each column can be made to generate a satisfying result of transforming after splitting the slice into numbers of horizontal columns. There are basically two strategies to realize the transformation, vertical shifting and rotating.

Vertical shifting refers to the strategy to transform the neuron network by shifting neurons in each column vertically to make a satisfying result of transformation. For neurons with different column indices, the magnitude of shifting may vary so as to achieve needed transformation. The vertical location of a soma position of a neuron is measured by its z coordinate in the study.

On the other hand, rotating involves rotation in the process of transformation. For every column, there are two different transforming matrixes available, which produce two different kinds of shapes after transformation --- shearing rotation and non-shearing rotation (see Fig. 6; Fig. 7). The non-shearing rotation adopts a normal rotation matrix (Harrington S., 1987) for each neuron in the same column. Both algorithms can alter the performance of the transformation by adjusting the position of the rotation centre and the rotation angle of each column.

Of note, a special non-shearing rotating method is employed to create the juxtaposed transformation (see Fig. 6). In this case, not only the neurons rotate but also the rotation centres rotate. For each loop in the algorithm, originally the bottom left corner of the rotation block. Then in the next ensuing loop, as the counter has increased, one piece of column is excluded from the rotation block of neurons and a same rotation event happens for neurons whose column index is equal or greater than the counter in the loop (which forms a rotation block) rotate by the transformed centre which is **Fig. 3.** Schematic diagram for the work flow of this project.

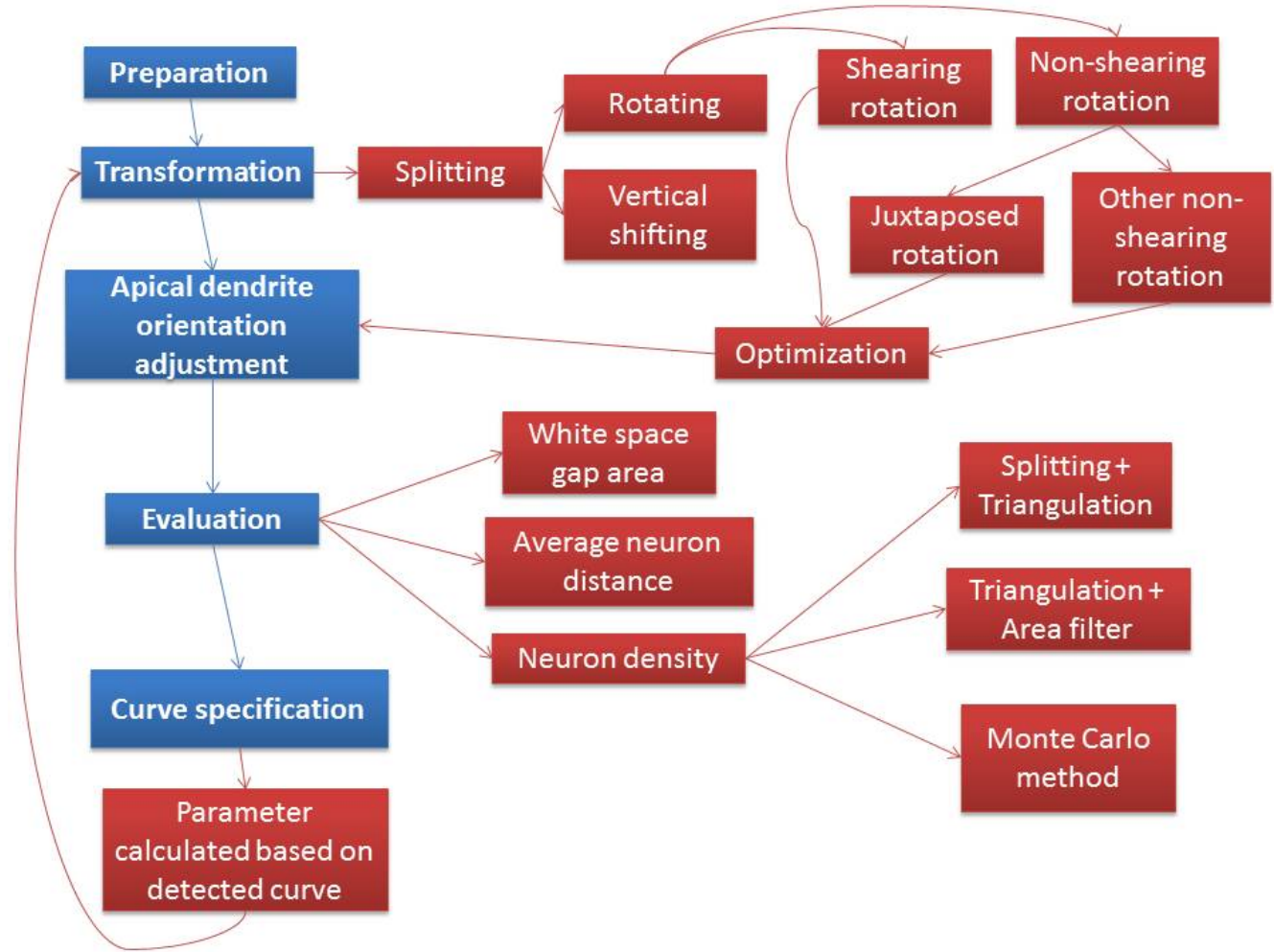

the updated rotation block with the updated rotation centre. In short, greater the index of the column is, more times of rotation the column encounters.

*2.2.3 Optimization* The aim of optimization is to undermine the unrealistic effects caused by transformation, such as the white spaces between columns after rotation, so as to better simulate neuron distribution in curved cortical layers in reality. Two methods are used to achieve the goal. Expansion of neuron positions in each column is utilized first to reduce the white spaces. By dividing each column into certain number of vertical aligned rows, different expansion ratios are assigned for each row, basically complying the rule that neurons with greater z coordinate have larger expansion ratio as the white space between these neurons are more enormous. After optimization, the efficiency of this algorithm is evaluated by plotting the neuron density heat map using the Matlab function HeatMap (Eisen et al, 1998).

### 2.3 Apical dendrite orientation adjustment

As is known that the orientations of apical dendrites of pyramidal neurons are always perpendicular to the cortical surface, adjustment must be made after the transformation. As for the methods of rotation, the angle which the apical dendrite should rotate is the same as the angle that the column at which the neuron locates has rotated. Contrarily, the case is more complicated for the methods of vertical shifting. Before the adjustment, the curve fitting the transformed cortical slice surface is calculated. Then for each pyramidal neuron, rotation of its apical dendrite is made with the angle that makes it perpendicular to the tangent of the calculated curve above. To simplify the algorithm, a point on the curve whose connection with the target neuron has the shortest distance among that of all points on the curve is found. And the orientation of that connection is used as that of the apical dendrite as it can be proved in analytical geometry.

### 2.4 Evaluation

To evaluate the performance of the transformation methods, three important criteria are examined by creating algorithms to calculate them from the transformed neuron network.

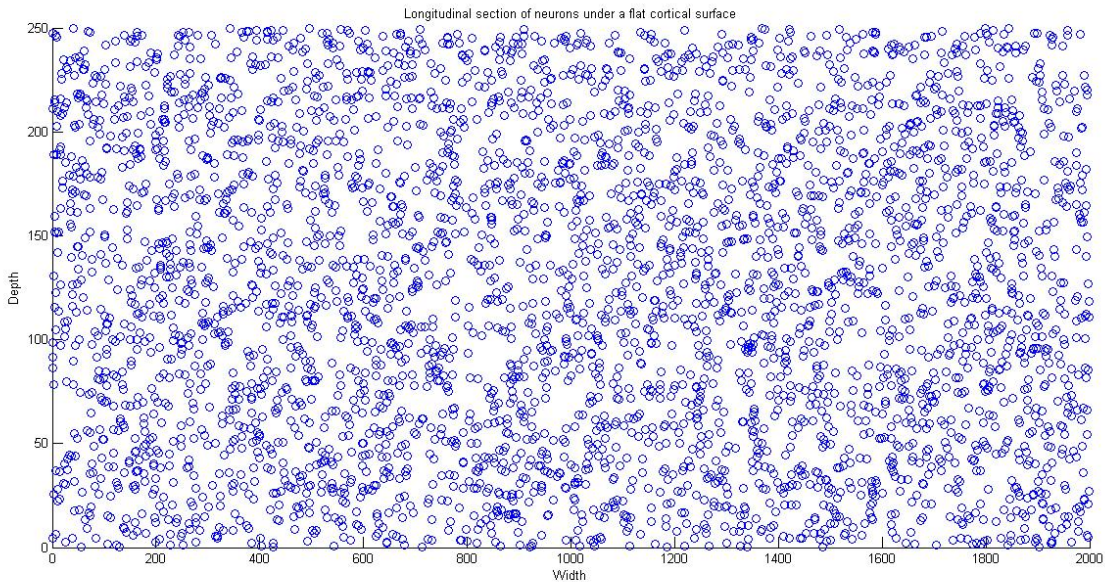

**Fig. 4.** Longitudinal section of neurons under a flat cortical surface. Vertical coordinate represents the depth of neurons and horizontal coordinate represents the width of neurons. All the following transformation displays are based on this view which is a longitudinal section.

*2.4.1 Neuron density evaluation* An essential criterion for evaluating the transformation methods is the change of neuron density. In order to calculate neuron density, the area where the neuron network covers should first be calculated. Three algorithms are designed for the purpose of estimating the area.

The first algorithm is splitting plus triangulation. At the beginning, the cortical slice is split horizontally as what the transformation has initially conducted. After that, neurons in the same column are connected to form many non-overlapping triangles using Matlab function Delaunay Triangulation (DT) (Shamos and Hoey, 1975; Lee and Schachter, 1980). Then the area of each triangle can be calculated and the sum of the area of all the triangles can be estimated as area of the cortical slice.

The second algorithm combines the function of DT (most importantly the function of ConvexHull (Graham and Ronald, 1972) which calculate the outer boundary of scattering data points) with an area filter. First, DT function goes through the whole cortical slice to triangulate the neuron network. After that, the ConvexHull function outlines the hull of the slice and all the neurons dropping on the convex stroke of the slice are identified. However, because the slice has certain extent of curvature and it is probably not in a fully convex shape, some area can be mistakenly regarded by the function as part of the slice. Therefore, a filter is needed to eliminate the additional triangles. If the area of the triangle is beyond a specific threshold, it is filtered out as not being part of the slice. To determine what the threshold should be, an experiment is conducted to uncover the safe range of threshold by investigating the relationship between calculated total area and the quantity of threshold. It is assumed that when the plot of the area against threshold reaches the first plateau, the corresponding range of threshold is the safe region of threshold.

The last algorithm to calculate the area of the slice is by using Monte Carlo method. To begin, curves that fit the top and bottom surface of the slice should be found. A curve function that best fits all of the neurons is found using polynomial regression of Fit function in Matlab. For finding the curve function for the top surface, the Fit function is used again for the neurons that lie above the formerly predicted curve. The loop of fitting updated neurons with curve function is then repeated till the number of data points is not sufficient to execute another round of fitting. In this way, the last predicted curve function is assumed to be the top surface curve function. Similarly, the bottom surface curve function can be found by employing the loop, updating the neurons by collecting those which locates below the formerly predicted curve. As the edge curves have been found, an experimental tablet which contains the entire slice can be created. Then, points can be virtually dropped in the tablet by creating a random matrix with specific size. In this study, the points that drop in the constraint area which is configured by the predicted top and bottom curved edges are considered to be in the status of IN. Consequently, those points that are not in the status of IN are considered to be in the status of OUT. The number of points with the status of IN are counted and the quotient between this number and the total number of points are calculated. Finally, area of the slice can be calculated by the quotient.

By the above three algorithms, area of the slice and the neuron density can be calculated.

*2.4.2 Average neuron distance evaluation* After neuron density measurement, average neuron distance is estimated to be the second factor evaluating the methods. For each neuron, its average distance with all other neurons is calculated. Then the average of the above calculated average distances is figured out.

*2.4.3 White space evaluation* The last factor used to evaluate the transforming methods is to estimate the area of white space in neuron gaps. As the factor is meaningless if the neurons are infinitely small, every neuron is assumed to occupy certain size of area within a specific radius, which means every neuron is a circle with the soma position being the centre. As a result, the value of radius is crucial determinant for the result of this algo-rithm. To find out an appropriate value for the radius which is neither too large nor too small, an experiment investigating the relationship between radius and the white space area of a flat cortical slice is launched. In this experiment, the value in search is the one where the white space area first drops to zero from the view of the plot. Only this condition is satisfied is the premise established that there is no white space in the untransformed cortical slice, which could be used as black group for the ensuing measurement for different transformation methods. With the radius being appropriately set, the above mentioned Monte Carlo method is used to estimate the area of gap space in the entire neuron network. By setting boundaries, distributing coordinates randomly in experimental tablet and counting number of coordinates in constraint area, the quotient and the gap space area are estimated.

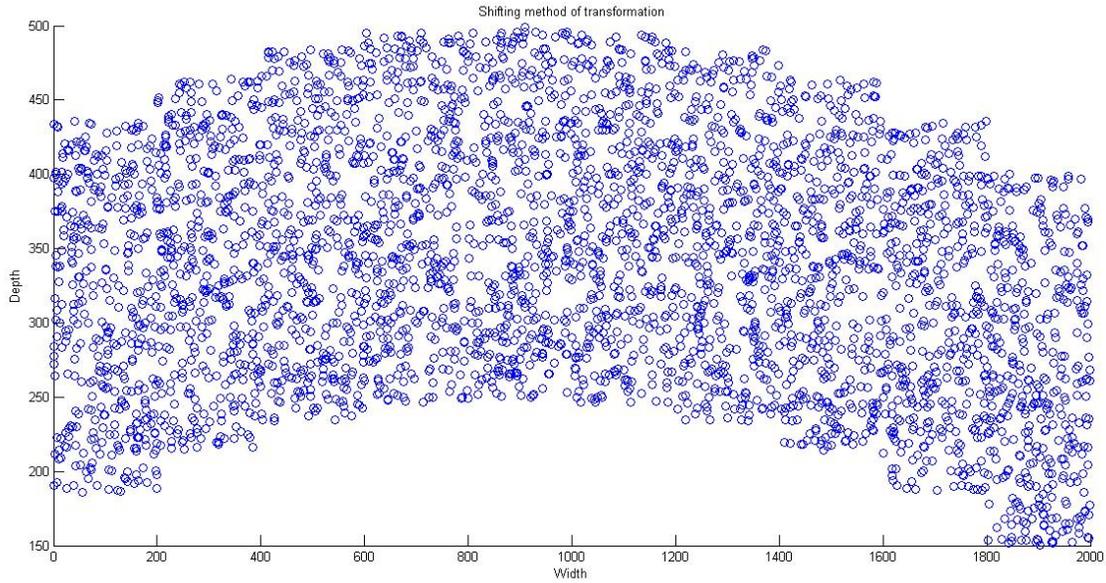

**Fig. 5.** Shifting method of transformation. The slice is imperceptibly divided into ten columns and each column shift vertically for an assigned distance to make the final curve output.

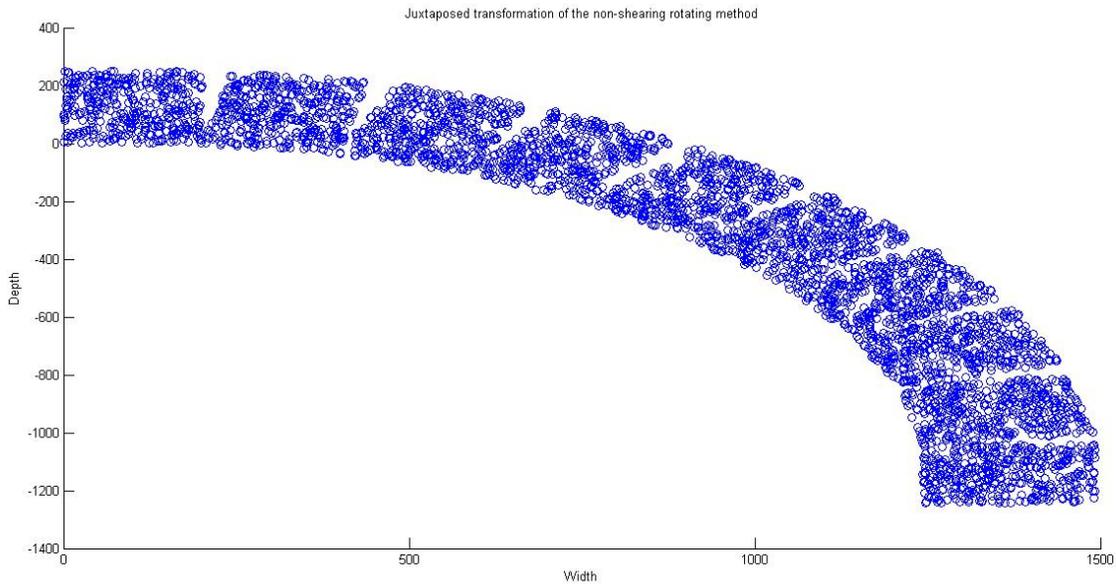

**Fig. 6.** Juxtaposed transformation of the non-shearing rotating method. In this case, the slice is divided into 10 columns. The bottoms of each column are concatenated though each of them rotates by an angle of a different degree.

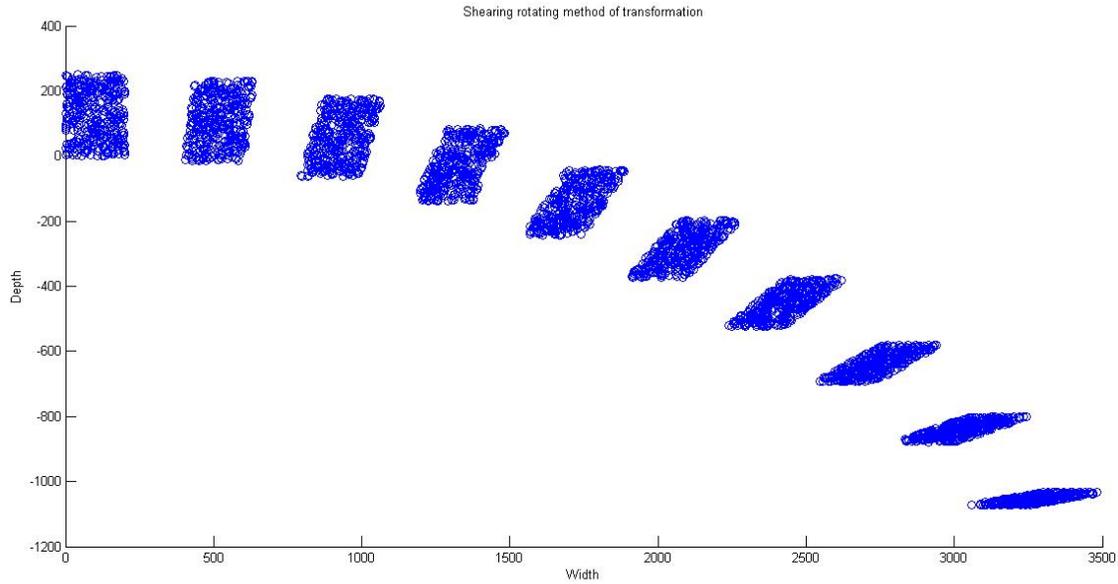

**Fig. 7.** Shearing rotating method of transformation. In this case, the slice is divided into 10 columns. Each columns are widely parted and gradually intensely sheared at horizontal direction.

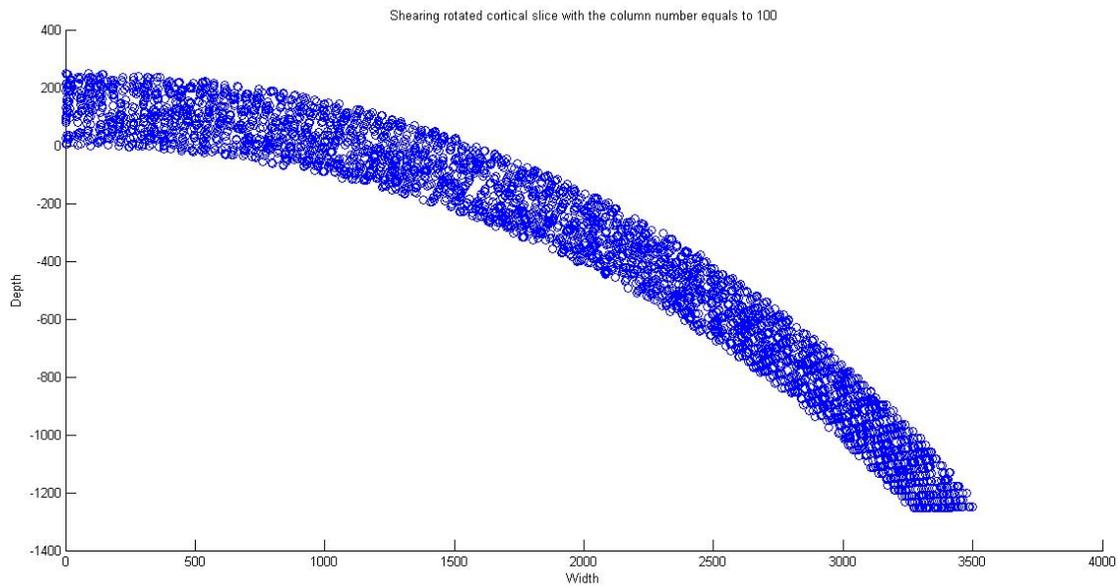

**Fig. 8.** Shearing rotated cortical slice with the column number equals to 100. Every transformation method can improve its performance by increasing the column number. It is because the edges of the slice become more continuous and smooth when there are more gradually changed columns to handle the transformation. It also excellently solves the problem of gaps between columns without optimization. Same rule applies for the non-shearing rotating method as it is a superior method compared to shearing rotating method. Under much less column number, for example, 50, it can preclude the unsatisfying jagged edges and smoothen the layout of the whole slice.

## 2.5 Curve specification

In order to fit the transformation to real experimental data, an algorithm is designed to make transformation according to detected curve information. As this part of realization is simple by means of a vertical shifting transformation method, only the algorithm for rotating method is described in this section. To initiate this algorithm, there is a prerequisite that the surface curve function of the experiment cortical slice should be detected through regression. If that function is revealed, the algorithm starts by setting the following parameters manually – the splitting number of horizontal columns, maximum and minimum column rotation angle, angle change type (linear or non-linear). Based on these assigned parameters, the positions of rotation centres are calculated. As all the parameters needed for a rotating transformation are known, transformation employing the methods introduced above can be initiated.

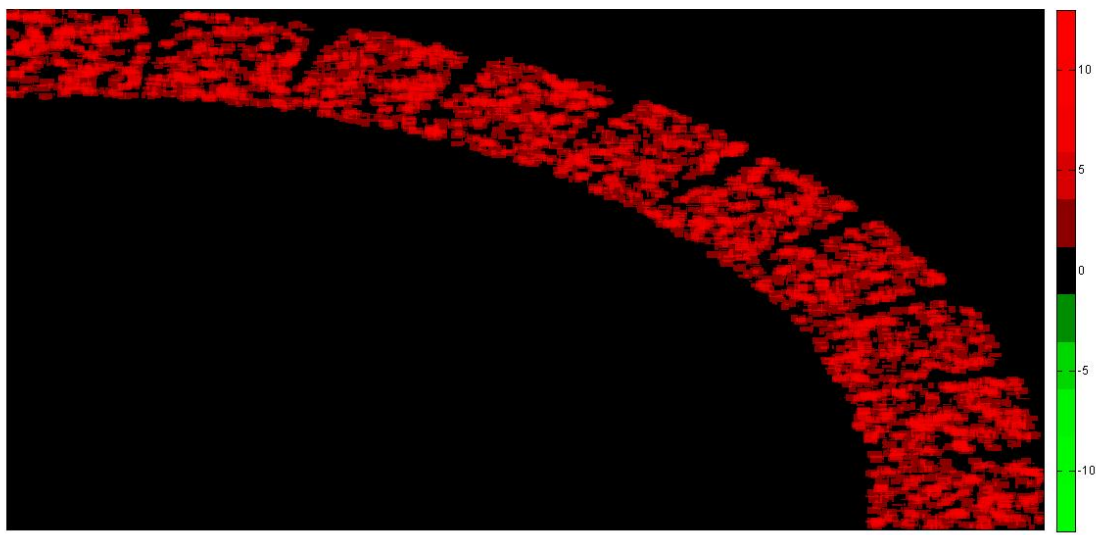

**Fig. 9.** The heat map of neuron density of neuron network by a juxtaposed transformation before optimization. More intense red pixel indicates denser neuron distribution in that area. Neurons are distributed obviously sparser in the gaps between columns after rotating transformation.

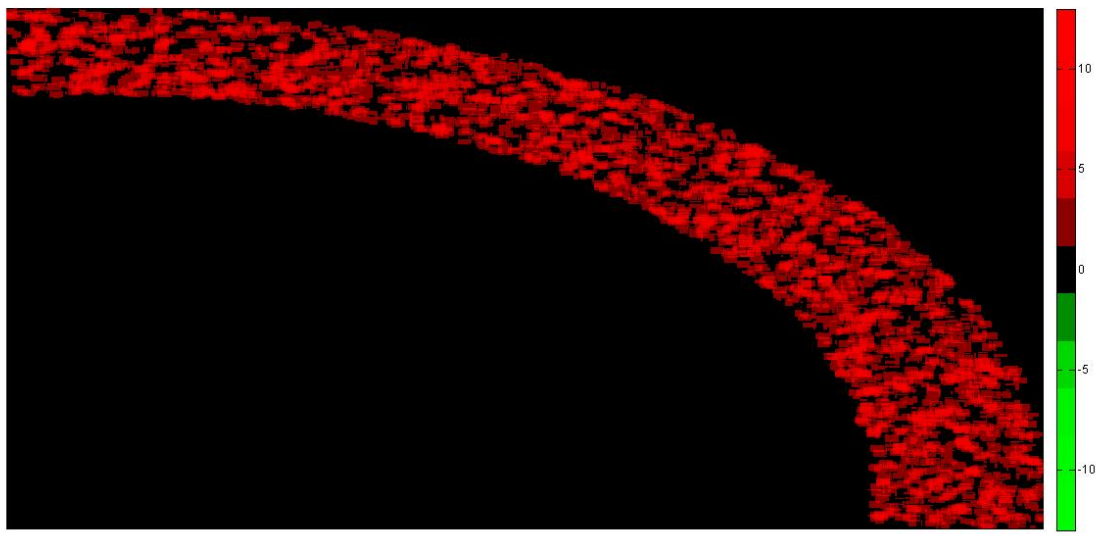

**Fig. 10.** The heat map of neuron density of neuron network by a juxtaposed transformation after expansion optimization. The density gaps disappear through the expansion optimization, which indicates the optimization is effective in eliminating column gaps generated by the process of transformation.

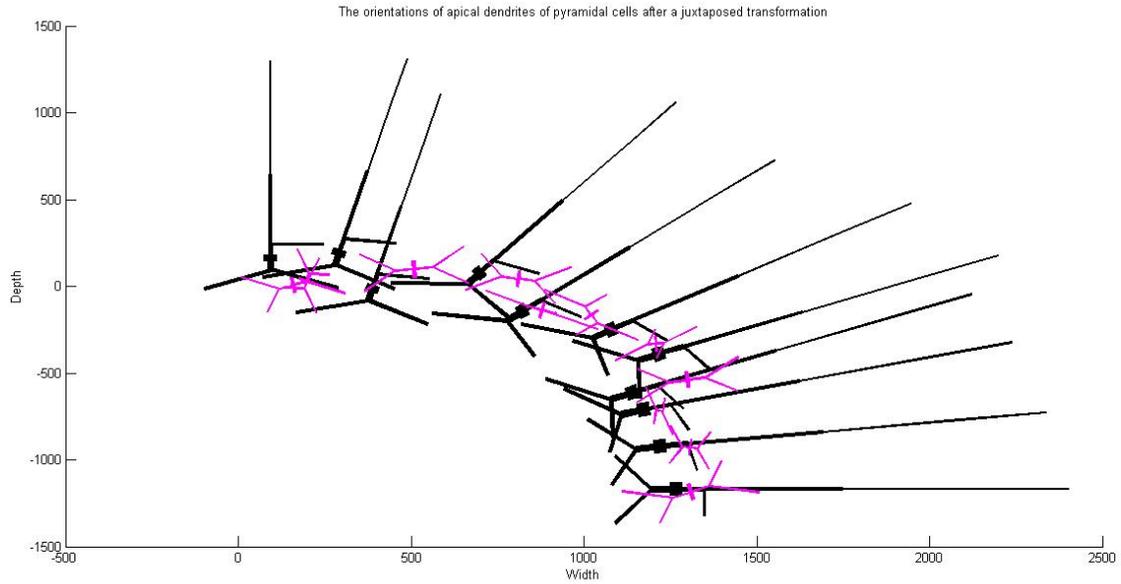

**Fig. 11.** The orientations of apical dendrites of pyramidal cells after a juxtaposed transformation. Note that the pink compartments refer to neurons that are not pyramidal neurons. These neurons remain to rotate randomly in a curved slice without obeying the rule of pointing toward to surface of the cortical slice.

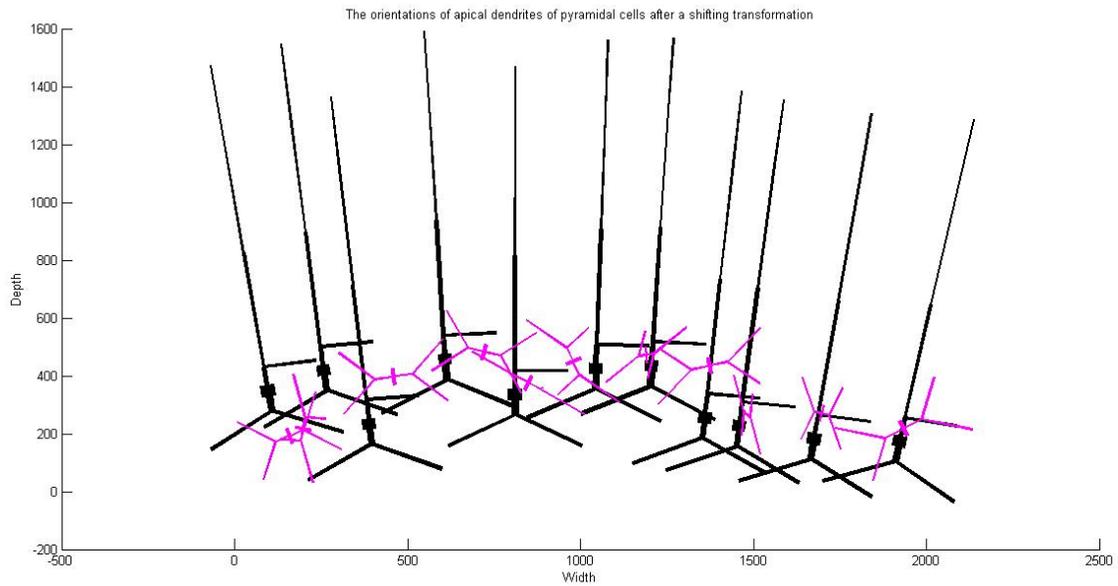

**Fig. 12.** The orientations of apical dendrites of pyramidal cells after a shifting transformation.

## 3 RESULTS

Prior to the transformation, simulation to prepare for a flat cortical slice is initiated and neuron position is plotted with x and z coordinates (see Fig. 4). Neurons are distributed in a rectangular area and scattering randomly.

Then vertical shifting and rotation methods are separately applied to achieve the transformation (see Fig. 5; Fig. 6; Fig. 7) all with the number of columns equals to ten. It appears that the vertical scale of transformed cortical slice from rotation is much larger than that from vertical shifting. Moreover, the slice transformed by rotation is more severely bent than that by vertical shifting. Of note, the comparison between shearing rotation and non-shearing rotation (especially juxtaposed position) reveals that transformation matrix of the latter method is superior to that of the former one. The reason is that neurons transformed by the former method are obviously parted resulting in extreme neuron density.

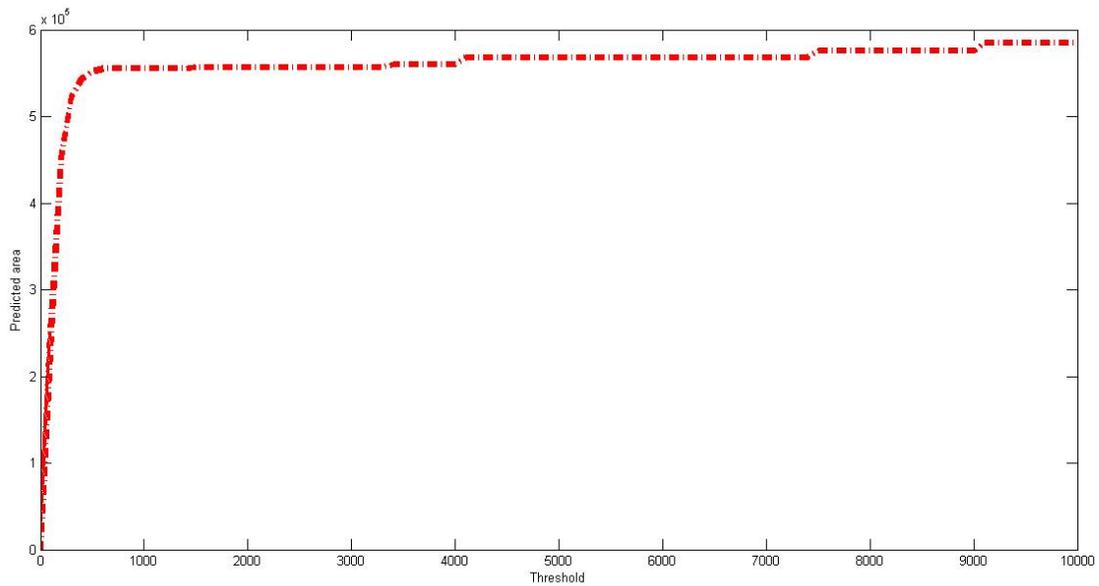

**Fig. 13.** Relationship between threshold (area of triangles) used to filter out large excessive triangles and the predicted entire slice area in the DT + area filter neuron density calculating method. Axis at the bottom represents the quantity of threshold. And the axis on the left represents the predicted entire slice area. The figure demonstrates that the predicted area mount at a nearly constant plateau when the value of threshold reach the range from 1000 to 4000. It is likely to be the vital threshold distinguishing the inherent triangles and the excessive triangles.

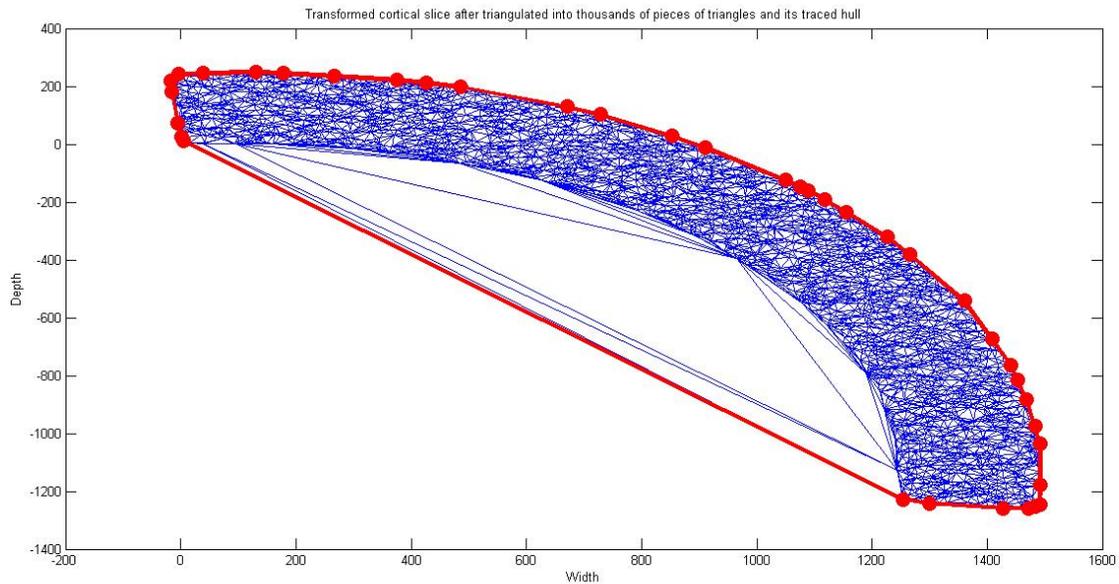

**Fig. 14.** Transformed cortical slice after triangulated into thousands of pieces of triangles and its traced hull. Red marker and line represent the convex hull of the slice. However, it is apparently erroneous that excessive triangles are included below the concave edge of the slice. A filter is needed to preclude these triangles. Note, the red line outlining the outer edge of the slice is the detected hull using function convexHull, within which the area of the slice is calculated.

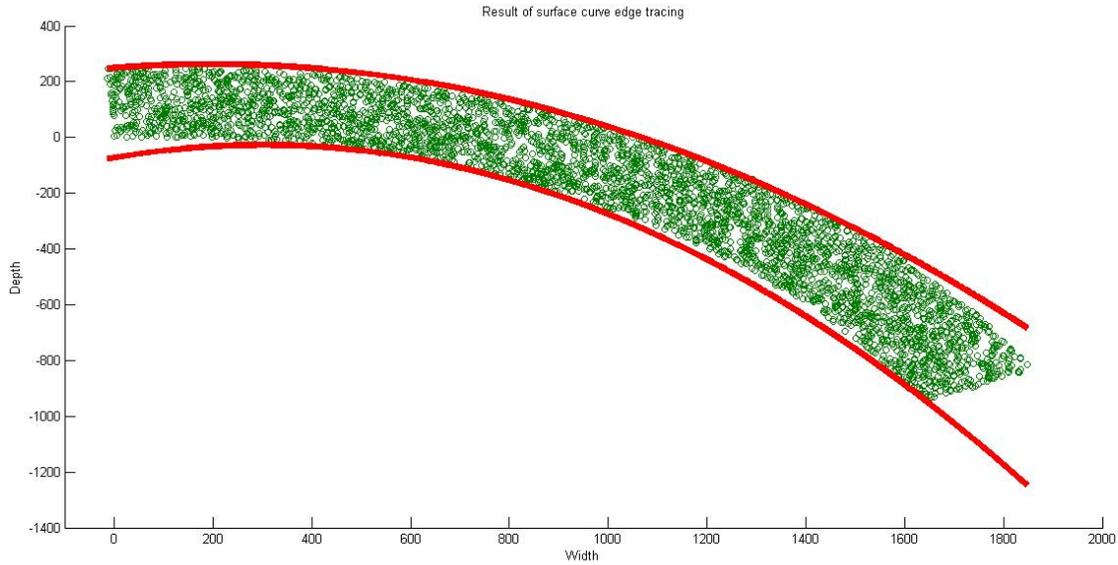

**Fig. 15.** Result of surface curve edge tracing. In the process of Monte Carlo neuron density calculating method, curve edges need to be traced and defined in order to execute following procedures. Recursive fitting is used for both the top and bottom edge to track the best fitting curve function. The figure shows that the algorithm works nicely in this case.

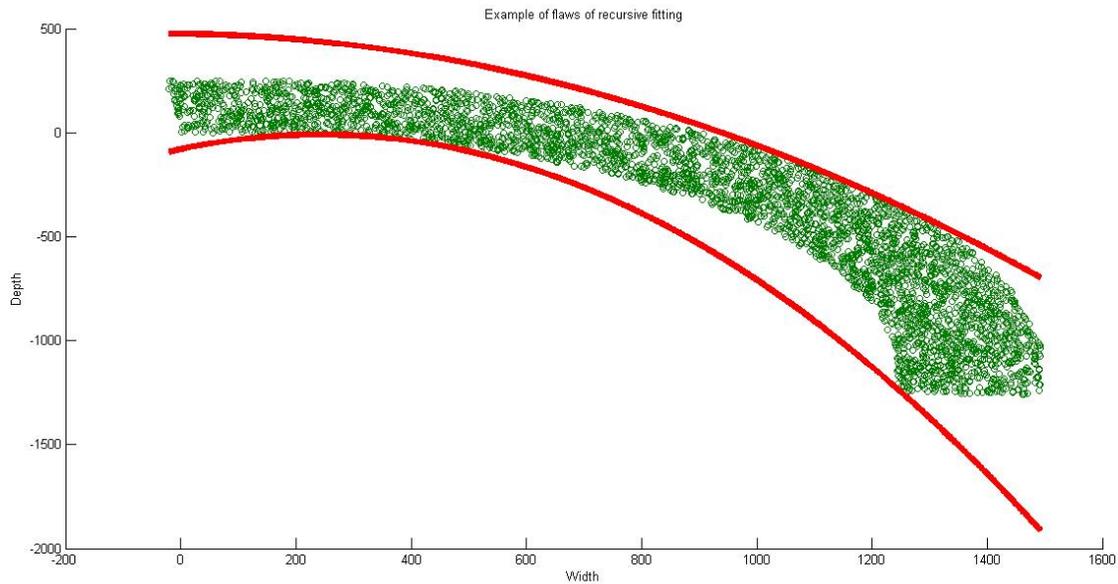

**Fig. 16.** Example of flaws of recursive fitting. The performance of recursive fitting degenerates when the slice is bent with a relatively large angle. It indicates that the algorithm is not robust enough and needs further optimization.

To compensate for the white space gap generated by the rotating transformation, an optimizing algorithm is designed to alleviate the visual deficit created by the gaps. Using the algorithm, it turns out that the scaling up with vertical gradient for each column successfully fills up the white space gaps (see Fig. 9; Fig. 10). Demonstrated by the figure, neuron density where the gaps reside originally returns to normal level in the optimized slice. And the result is satisfying as it greatly match the real configuration of cortical layers.

The adjustment of orientation of apical dendrites of pyramidal cells is also successful (see Fig. 12; Fig. 13). Every apical dendrite is facing toward the curved surface.

Evaluation of these transformation methods focus on three measurements - neuron density, white space area and average neuron distance. Worth noticing, slice area used to calculate neuron density is estimated separately through three different algorithms. The performance of the first method (Splitting + DT) appears to be intensely affected by the number of columns split before transformation. A remarkable trait of this algorithm is that the predicted area decrease as the number of columns increases (see Table. 1).

The area filter is the key to the second algorithm. To search for an appropriate threshold, experiment investigating the relationship between threshold and predicted slice area (see Fig. 13) illustrates that the increment of area slow down dramatically after threshold reaching 1000. Inferred from the remarkable area distinction between inherent triangles and excessive triangles (see Fig. 14), 1000 might be the starting point of a reasonable threshold range. And as a noticeable bounce is detected when threshold reach 4000, it is plausible to assign it to the end point of threshold range. This range is proved to be acceptable as the area remains constant (5.4940+e05) when threshold equals to 2000 and 3000 (see Table. 1).

**Table 1.** Area predicted by three neuron density algorithms (estimated value: 5.4909+e05)

| Methods | Area | | | |
|---|---|---|---|---|
| Splitting + DT | 5.5158+e05 ($N_c$ =5) | 5.3008+e05 ($N_c$=10) | 4.6358+e05 ($N_c$=50) | 4.1122+e05 ($N_c$=100) |
| DT + filter | 5.4673+e05 (thr=1e+03) | 5.4940+e05 (thr=2e+03) | 5.4940+e05 (thr=3e+03) | |
| Monte Carlo | 5.2963+e05 ($N_p$=e+04) | 5.1489+e05 ($N_p$=e+05) | 5.1877+e05 ($N_p$=e+06) | 5.1807+e05 ($N_p$=e+07) |

$N_c$ = Number of columns split before calculation; thr = Threshold of triangle area filter; $N_p$ = Number of points employed in Monte Carlo experiment.

The third area prediction method developed employs the famous Monte Carlo method. In this strategy, it is vital to accurately trace the edge of the entire cortical slice. The recursive fitting algorithm is proved to function perfectly in most cases as expected (see Fig. 15). However, erroneous fitting outcome occur in some extreme cases in spite of the seeming robust algorithm (see Fig. 16).

As for the white space calculation algorithm, it is important to determine an appropriate radius that is used to define the constraint area in Monte Carlo experiment. In search for that value, the result of the blank test (see Fig. 17) implies that the white space area is approaching zero when value of radius increase to 15 or above. In other tests that are not presented, this watershed of value could be from 20 to 25. Therefore, a selected radius ranging from 15 to 25 is considered to be a reasonable choice to define the constraint area in Monte Carlo experiment of this calculation.

Finally, the performance of algorithm to transform the flat slice into a curve slice fitting specific function is evaluated. Although the predicted curves are roughly aligned to both the top and bottom surface, quite huge mismatches are found from the plot of the result (see Fig. 18). The discrepancy between quadratic coefficients clearly demonstrates the existence of mismatches as the shape and curvature of the parabola is solely affected by this coefficient (see text in Fig. 18). Despite its imperfection in some cases, this algorithm produces excellent output if the absolute value of the quadratic coefficient is small enough, preferably less than 0.00003 with the presented settings of maximum and minimum angle. In this condition, not only the transformation algorithm but also the edge finding algorithm has good performance (see Fig. 19).

## 4 DISCUSSION

The key of neuron density calculation resides in the prediction of cortical slice area. In this study, three algorithms to calculate slice area are developed. The result (see Table. 1) shows various performance and reliability for different algorithms.

**Table 2.** Evaluation of transformation methods by neuron density, white space area and neuron average distance

| Methods | Neuron density | White space (R= 25) | Average neuron distance |
|---|---|---|---|
| Flat surface | 0.0077 | 0 | 681.5697 |
| Shifting | 0.0074 | 1.9285+e03 | 682.733 |
| Rotating | 0.0071 | 5.7697+e04 | 720.76 |

R = Radius, refers to the size which each neuron occupies in Monte Carlo experiment employed in calculating white space area.

Huge fluctuation of predicted value is found in the first algorithm. One reason is that more inherent triangles formed between neurons in adjacent columns are ignored due to the rise of column number. The other reason is more neurons are ignored because triangulation requires at least three neurons and larger number of columns makes less neuron at each column. Therefore, this method is a column-number-dependent method and lacks stability which can be perceived by the result.

Regarding the second method (DT + filter), the selection of threshold value is crucial because an unreasonably large threshold will not filter out excessive triangles (see Fig. 14) and an unreasonably small threshold will filter out inherent triangles which should be counted in the area. The result indicates that the proximity between predicted value and estimated value (5.4909+e05) implies that the method is accurate and reliable.

For the third method using Monte Carlo to estimate the area, the reason for the deficiency is that in some cases there exists tangent on the curved surface that is perpendicular to the x axis, which can be caused by a 90 degree rotation transformation of a particular column. As a function can only divide coordinate points into two parts as the above part and the below part, dividing them into left-hand part and right-hand part is impossible for any function. Consequently, error happens when the algorithm cannot correctly select the target neuron group for a next curve fitting. So, further study can be conducted to optimize the neuron updating algorithm at each loop of fitting in this method. Another case with erroneous performance happens when columns are separated widely apart from each other. It could have made devastating mistakes at the first round of curve fitting. Besides, improvement can be made to enhance the user experience by changing the FOR loop into a WHILE loop in the process of recursive fitting. This is because fit function needs at least three data points to determine three coefficients while how many loops are needed for the completion of fitting is not known beforehand. As a result, a WHILE loop instead of a FOR loop should be employed in this case.

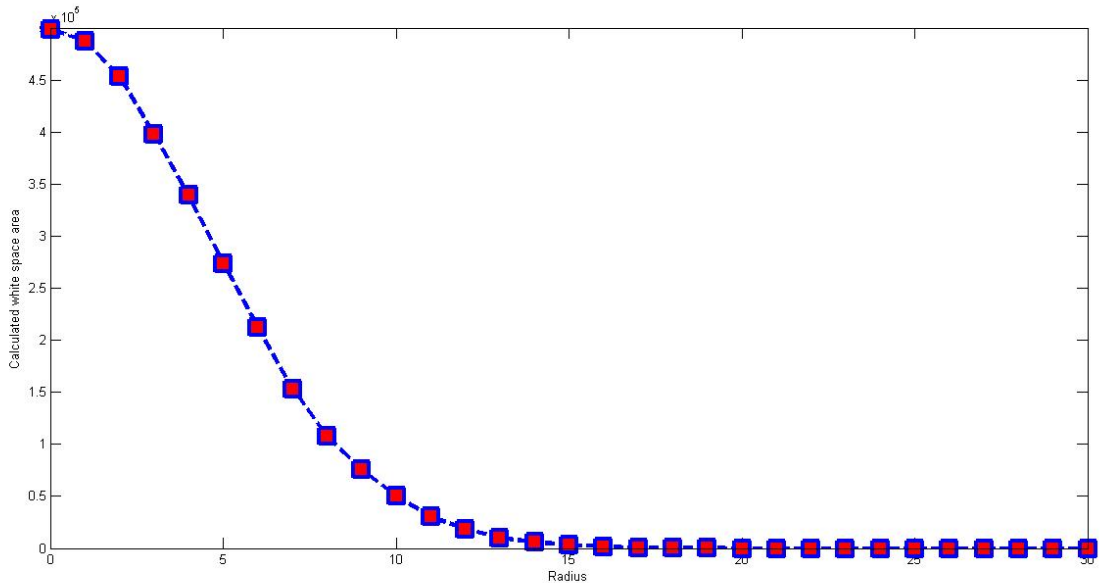

**Fig. 17.** The relationship between radius set in Monte Carlo experiment in white space area estimation algorithm and the estimated white space area of original flat cortical slice. This figure indicates that the white area of the original flat cortical slice drops to zero when the radius reaches or is beyond 20. Thus, 20 may be an appropriate radius for the boundary setting in the Monte Carlo experiment to calculate the white space area in transformed cortical slices. By exploiting this value, the premise is constructed that there is no white space in the untransformed cortical slice so that comparison can be made further between different transformed cortical slices from different algorithms.

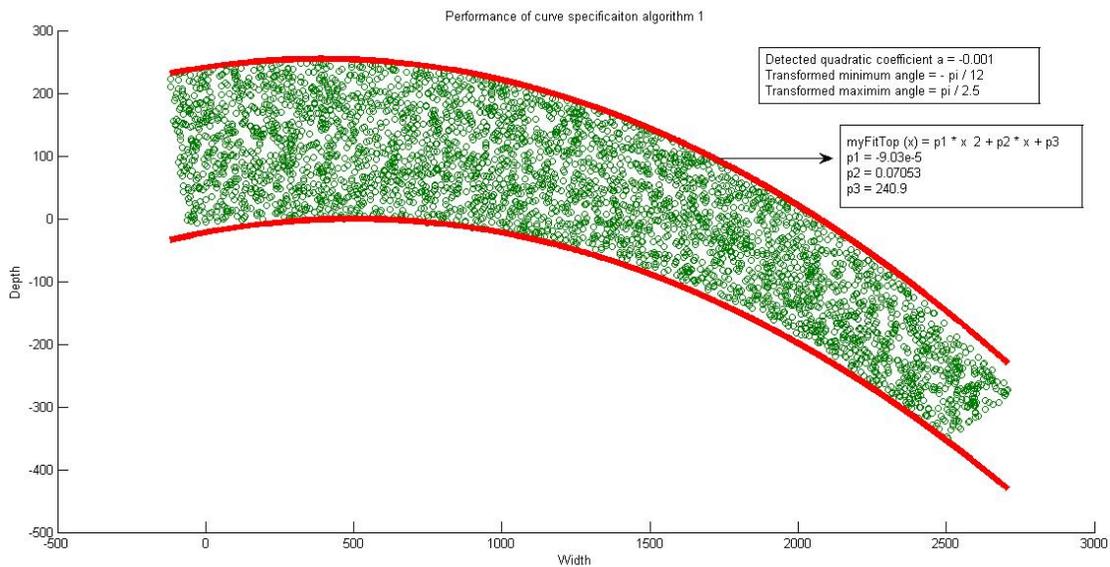

**Fig. 18.** The performance of curve specification algorithm. It is assumed that the quadratic coefficient of the detected curved function for the cortical surface is –e-03. After employing the algorithm, the surface of cortical slice turns out to fit a function with a quadratic coefficient of -9.03*e-05, by the curve finding algorithm described in method section. It is quite huge a deficit with more than 10 times disparity.

Basically there are two transformation methods developed in this study. Both shifting and rotating method have strengths and weaknesses. As for shifting method, the transformed cortical slice maintains excellent uniformity with the original flat cortical slice, which can be perceived by the three evaluation measurements (see Table. 2). The average neuron distance of shifting transformed cortical slice is around 682, exceeding just about 1 unit compared to that of the original slice. In addition, it is easy to manipulate the parameters in this method so as to achieve specified transformation, for example, a cortical slice whose surface fits a specific quadratic function, because the detailed settings to initiate the transformation in this method is based on given function. On the other hand, it cannot transform a flat cortical slice into an extensively curved slice, for instance, a slice bent over 90 degrees. On the contrary, rotating transformation methods display good flexibility with no limitation on the extent to which the slice is bent. It can handle however large the angle is. But, the weakness of these methods is also obvious that the transformed slice lacks uniformity and consistency with the original flat slice. The comparison of average neuron distance and white space area has shown that the entire slice is expanded and lots of extra space is added which prominently decrease the neuron density. Nevertheless, the performance of both the two methods can be improved by increasing the number of columns so that a smooth curved surface can be constructed (see Fig. 7; Fig. 8).

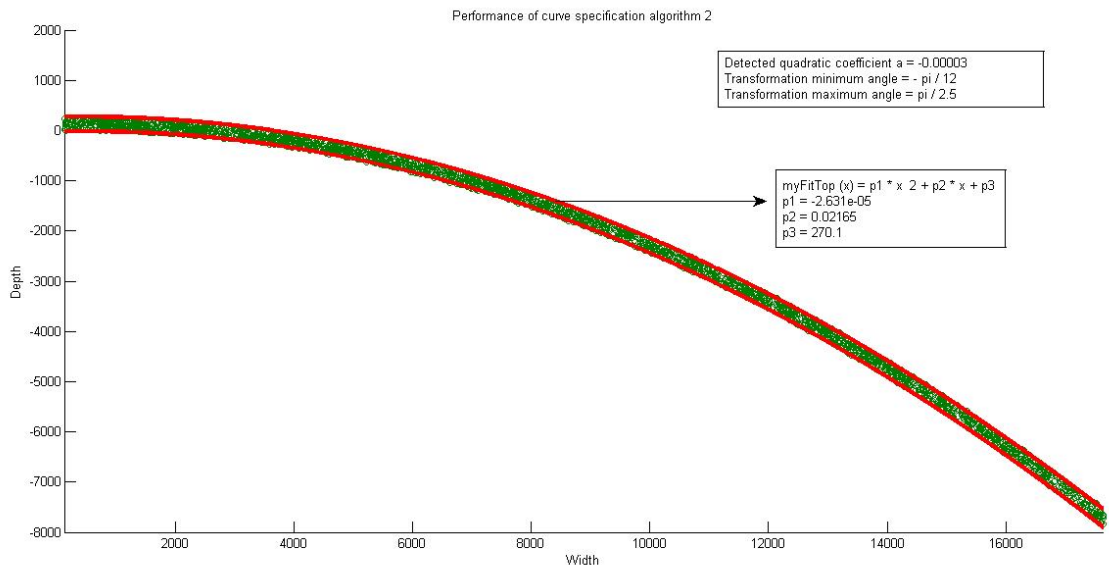

**Fig. 19.** Better performance of curve specification algorithm. The assumed detected absolute value of quadratic coefficient is much smaller (in this case, -3*e-05). As the figure indicates that the quadratic coefficient of the predicted transformed cortical surface curve (-2.631*e-05) is much closer to the assumingly detected one compared to the huge deficit in the previous case. It proves that the curve specification algorithm can perform nicely if the quadratic coefficient is small enough.

Transformation matrix is an important factor affecting the outcome of transformation. Comparing shearing rotation with non-shearing rotation, the quality of both methods can be revealed that cortical slice transformed by non-shearing rotation is better with less space occupied and is more consistent with the original flat slice. Different transformation matrixes being employed contribute to the different performance. The fundamental discrepancy of the two algorithms stems from the different settings of the rotation centre. As regard to non-shearing rotation, there is only one rotation centre for neurons of each column. All neurons in a particular column rotate equal amount of degree by a same centre. Markedly differently, shearing rotation employs multiple rotation centres for neurons in one column based on their x coordinates.

The tremendous deficit in the curve specification algorithm can be caused by two possible reasons. First of all, the problem may stem from the imperfection of the transformation algorithm that is unable to modify the cortical slice as it is demanded. Another probable reason for this is that the predicted surface curve functions are not exactly the real ones covering the top and bottom of the slice. As it has been discussed above that the edge finding algorithm developed from recursive fitting has fatal leaks and it may need further improvement and optimization to perform perfectly. Furthermore, the preferable limitation of quadratic coefficient (<0.00003) for the algorithm may results from the fact that all the parameters assigned in the algorithm need to be in harmony. For instance, if the quadratic coefficient is too small, resulting in so narrow a space that it cannot even withhold the entire flat slice, subsequent alternations or unexpected errors will happen leading to bad outcome. Moreover, the question that whether some parameters in the algorithm (for example, the maximum angle and minimum angle for columns) should be maintained also needs further discussion and investigation. In all, improvement and optimization is needed for this algorithm and an overall reconstruction of the cortical slice to simulate a curved surface is also a pragmatic choice for this approach.

Further extending and optimizing the transformation methods may include considering in the changes of layer thickness and neuron number in different layers. As is reported that both the relative thickness and absolute thickness of layers tend to alter under some patterns and there might exist neuron migration possibilities (Hilgetag and Barbas, 2006) that could greatly influence the number of neurons in folded and curved cortical tissue. By adopting these anatomical, physiological and mechanical characteristics from previous researches, more realistic and accurate algorithms simulating the neuronal activities of curved cortical tissue could be developed.

## ACKNOWLEDGEMENTS

I would like to give the utmost appreciation to my supervisor Marcus Kaiser for all the guidance, tuition and suggestion he had offered me about this project. The project could not have reached this far without his help. Also, I would like to gratefully thank Richard Tomsett who always intuitively and patiently gave technical help and support to my project from proposal phase to code scripting period. In addition, those anonymous online helpers who create and answer the threads relevant to the problems associated with my project are also coauthors of this article. I have got lots of assist from them and I am thankful to all of them. Moreover, I recognize the developers who have innovated clever and useful algorithms such as Delaunay Triangulation etc. At last, great acknowledgement is given to the Internet for all the online resources and literatures.

## APPENDIX

Software: Matlab
Version: R2012a (7.14.0.739)
Operating system: 64-bit (win64)
Machine configuration
Processor: Intel® Core ™ i5-2450M CPU @ 2.50Hz 2.50Hz
Installed memory (RAM): 4.00 GB